\documentclass[twocolumn,aps,prb,superscriptaddress]{revtex4-2}
\usepackage{graphicx,bm,natbib}
\usepackage{amssymb}
\usepackage{amsthm}
\usepackage{amsmath}
\UseRawInputEncoding
\usepackage{color}
\usepackage{slashed}
\usepackage{float}
\usepackage{tikz}
\usepackage{pgf}
\usepackage{pgfplots}
\usepackage[hidelinks]{hyperref}
\hypersetup{
    colorlinks=false,
    }

\usepackage{amsmath,amsfonts,amssymb}
\usepackage{braket}

\newcommand{\be}{\begin{equation}}
	\newcommand{\ee}{\end{equation}}

\newcommand{\bea}{\begin{eqnarray}}
	\newcommand{\eea}{\end{eqnarray}}

\newcommand{\p}{\partial}

\newcommand{\lp}{\left(}
\newcommand{\rp}{\right)}

\renewcommand{\vec}[1]{{\boldsymbol #1}}
\newcommand{\ba}{\begin{array}}
  \newcommand{\ea}{\end{array}}

\makeatletter
\let\Hy@backout\@gobble
\makeatother

\begin{document}
\title{
Hofstadter quasicrystals, hidden symmetries 
and irrational quantum oscillations} 

\author{Kirill Kozlov}
\affiliation{Moscow Institute of Physics and Technology, Dolgoprudny, 141700 Russia}
\author{Grigor Adamyan}
\affiliation{Department of Physics and Astronomy, Johns Hopkins University, Baltimore, MD 21218, USA}
\author{Mariia Kryvoruchko}
\affiliation{Department of Biophysics, Johns Hopkins University, Baltimore, MD 21218, USA}
\author{Yelizaveta 
Kulynych}
\affiliation{Department of Physics, Taras Shevchenko National University of Kyiv, Kyiv 03680, Ukraine}
\affiliation{Department of Physics, The \'Ecole polytechnique f\'ed\'erale de Lausanne, Lausanne 1015, Switzerland}
\author{Leonid Levitov}
\affiliation{Physics Department, Massachusetts Institute of Technology, Cambridge MA02139, USA}


\begin{abstract}
Landau levels perturbed by a periodic potential is a prime setting to design quantum systems with exotic fractal spectra. Motivated by recent advances in twistronics, we introduce `Hofstadter quasicrystal' problem describing Landau levels perturbed by a set of incommensurate cosine waves. We illustrate the underlying physics for moir\'{e} quasicrystals with octagonal and dodecagonal symmetries, finding spectra that are vastly more complex than the Hofstadter spectrum. Surprisingly, due to the high spatial symmetry, the quasicrystal problem exhibits hidden `inner' symmetry arising at special `magic' values of the magnetic field. The $1/B$-periodic pattern of magic field values explains striking wide-range oscillations in the observed spectra that have irrational periodicity incommensurate with the Aharonov-Bohm and Brown-Zak periodicities. The prominent character of these oscillations makes them readily accessible in state-of-the-art moir\'{e} graphene systems. 
%
\end{abstract}

\maketitle

Perturbing Landau levels by periodic potentials offers seminal examples of quantum fractals \cite{Satija,Hofstadter_1976,
Thouless,Wannier_1978,MacDonald1983}. The intricacy, richness, and general character of the fractal behavior originate from the unique symmetries 
governed by magnetic translations. These are symmetry operations  
that commute with the Hamiltonian but not with each other, and, together, span a large symmetry group \cite{Brown1964,Zak1964}.
Magnetic translations make the phases of particle states depend on the order in which translations are performed. This results in the disappearance and recurrence of Bloch bands, governed by commensurability of the magnetic flux through the lattice period, 
generating self-similar fractal bands resembling a Cantor set. This intriguing behavior has garnered significant interest over the years and has been extensively studied for the case of Landau levels of Bloch electrons  
\cite{
Thouless,Thouless1982,Rammal1985,Wannier_1978,MacDonald1983,
Wiegmann1998,Avron_2003,Kohmoto2006,Roy2016}, 
\begin{align}\label{eq:H_general}
    H=\epsilon(\vec p-e\vec A(r))+V(\vec r)
    ,
\end{align}
where $V(\vec r)$ is a periodic potential, $\epsilon(\vec p)$ is particle kinetic energy and vector potential $\vec A(r)$ describes a uniform magnetic field. 
Exquisite number-theoretic and topological structures discovered for this problem 
suggest extensions to a wider family of non-periodic potentials. Recently, a number of generalizations to hierarchical and aperiodic lattices have been considered 
\cite{Nandy2014,Fuchs2018,Marques2023}. At the same time, magnetotransport measurements unveiled fractal behavior in graphene superlattices \cite{Dean_2013,Ponomarenko_2013,Hunt_2013,BrownZak_Manchester_2017, BrownZak_Manchester_2018,Lu_2021}. 

\begin{figure}[t]
    \centering
    \includegraphics[scale = 8.10]{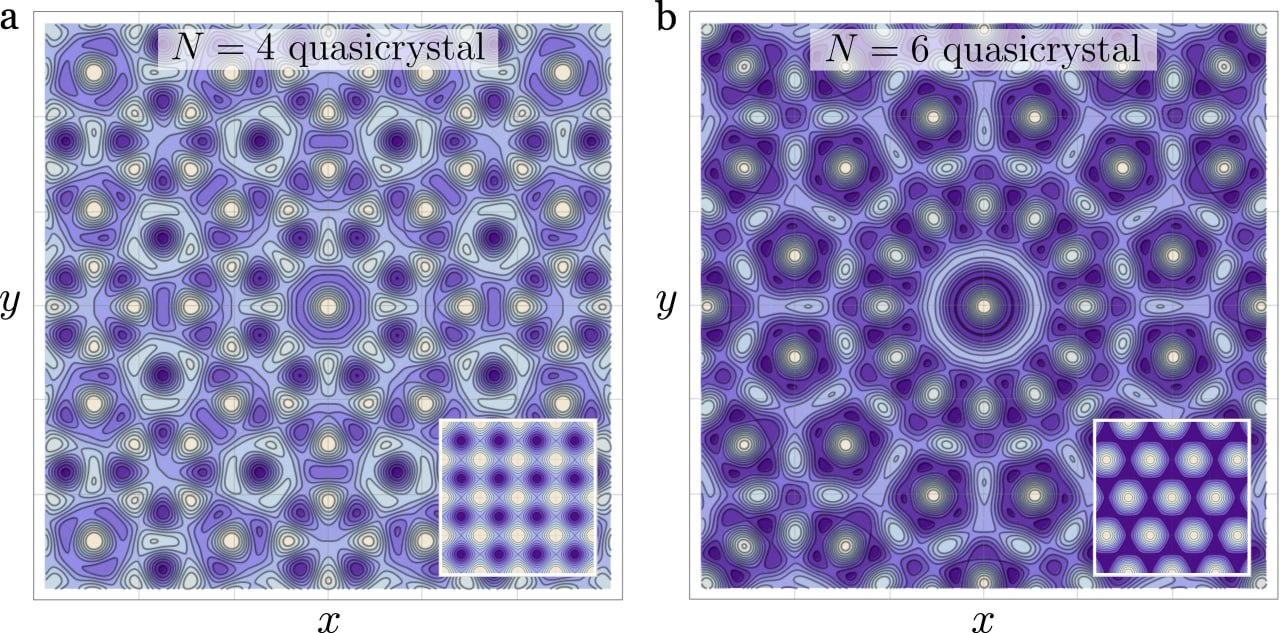} 
    \caption{
    Octagonal and dodecagonal moir\'e quasicrystal potentials, 
    Eq.\,\eqref{eq:H_incommensurate}. The problems possess
8-fold and 12-fold 
    point symmetry groups comprising rotations and mirrors. Analysis of the quantum problem described below relies on representing these potentials as pairs of the periodic potentials shown in the insets rotated by $45^{\circ}$ and $30^{\circ}$, respectively. 
    }
    \vspace{-5mm}
    \label{fig:PS}
\end{figure}

\begin{figure*}[t]
    \centering
    \includegraphics[scale = 1.85]{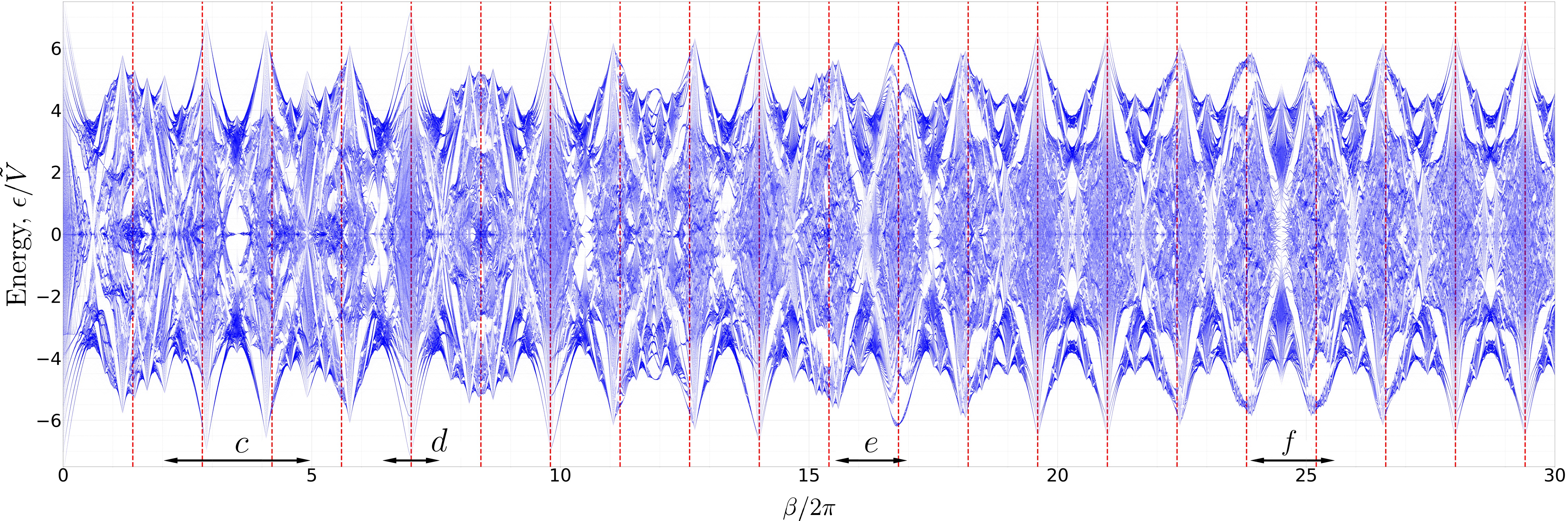}
    \caption{Energy spectrum of an octagonal quasicrystal, Eq.\eqref{eq:H0+H1},  
    vs.  $\beta$ (the inverse $B$ field) obtained from a Harper-like 1D problem. The factor $\tilde V$ describing the oscillations decay with $\beta$, Eq.\,\eqref{eq:Debye_Waller_factor}, is factored 
    in the energy units. Note pairs of positive and negative pyramids filled with Landau-like fans of energy states and complex deep-level structure of spectral gaps detailed in Fig.~\ref{fig:SKB_spectrum}, where zoom-ins of the regions marked by c), d), e) and f) are shown. The modulation in the spectrum width seen on a large scale is explained by a recurring pattern of hidden symmetries originating from magnetic translations, see Eq.\,\eqref{eq:H1H2_commutator}. Red lines mark the 'magic' $\beta$ values defined in Eqs.~\eqref{eq:irrational_betas}, at which the $N=4$ quasicrystal Hamiltonian, Eq.~\eqref{eq:H0+H1}, decouples into a pair of commuting Hofstadter $N=2$ Hamiltonians,  Eq.~\eqref{eq:H_2cos}. This irrational periodicity, appearing over a wide range of fields and not explained by the Aharonov-Bohm or Brown-Zak oscillations represents a readily observable signature of the Hofstadter quasicrystal. 
    }
    \vspace{-4mm}
    \label{fig:SKB_spectrum_lines}
\end{figure*}

However, relatively little is known about the more general problem of Landau levels perturbed by potentials with incommensurate periods. Here we consider moir\'e quasicrystal potentials comprising harmonics defined by stars of vectors with non-crystallographic symmetries: 
\begin{align}\label{eq:H_incommensurate}
    & V_N(\vec r)=
    2V\cos \vec k_{1}\vec r
    +...+
    2V\cos \vec k_{N}\vec r
    ,
    \\ \nonumber
    & \vec k_{s}=(\kappa \cos\theta_s, \kappa \sin\theta_s)
    ,\quad
    \theta_{s=1...N}=\pi s/N 
    ,
\end{align}
which defines a crystal for $N=2,\,3$ and a quasicrystal for $N>3$ 
(see Fig.\,\ref{fig:PS}) \cite{quasicrystal_vs_moire}.
On a Landau level, this defines a quantum problem for noncommuting coordinates of a cyclotron orbit center, see below. Besides an obvious fundamental interest, this problem is timely due to the advent of twistronics --- a field that produces and probes incommensurate 2D systems comprising two or more periodic lattices rotated relative to one another by a fixed angle and strongly coupled at the atomic scale
\cite{Carr2017,Bistritzer,Cao2018}. 
Octagonal quasicrystals obtained from $45^{\circ}$-twisted stacks of high-$T_c$ superconductors have been widely studied as a platform for topological superconductivity 
\cite{Vishwanath2021,Can2021,Volkov2020,Lee2021,Zhao2021,Glazman2022,Song2022}. Likewise, dodecagonal quasicrystals are readily available in state-of-the-art moir\'e graphene toolbox 
\cite{Ahn2018,Yao2028,Spurrier2019,Moon2019,Yu2019,Pezzini2020}.


\begin{figure*}[th]
    \centering
    \includegraphics[scale = 1.98]
    {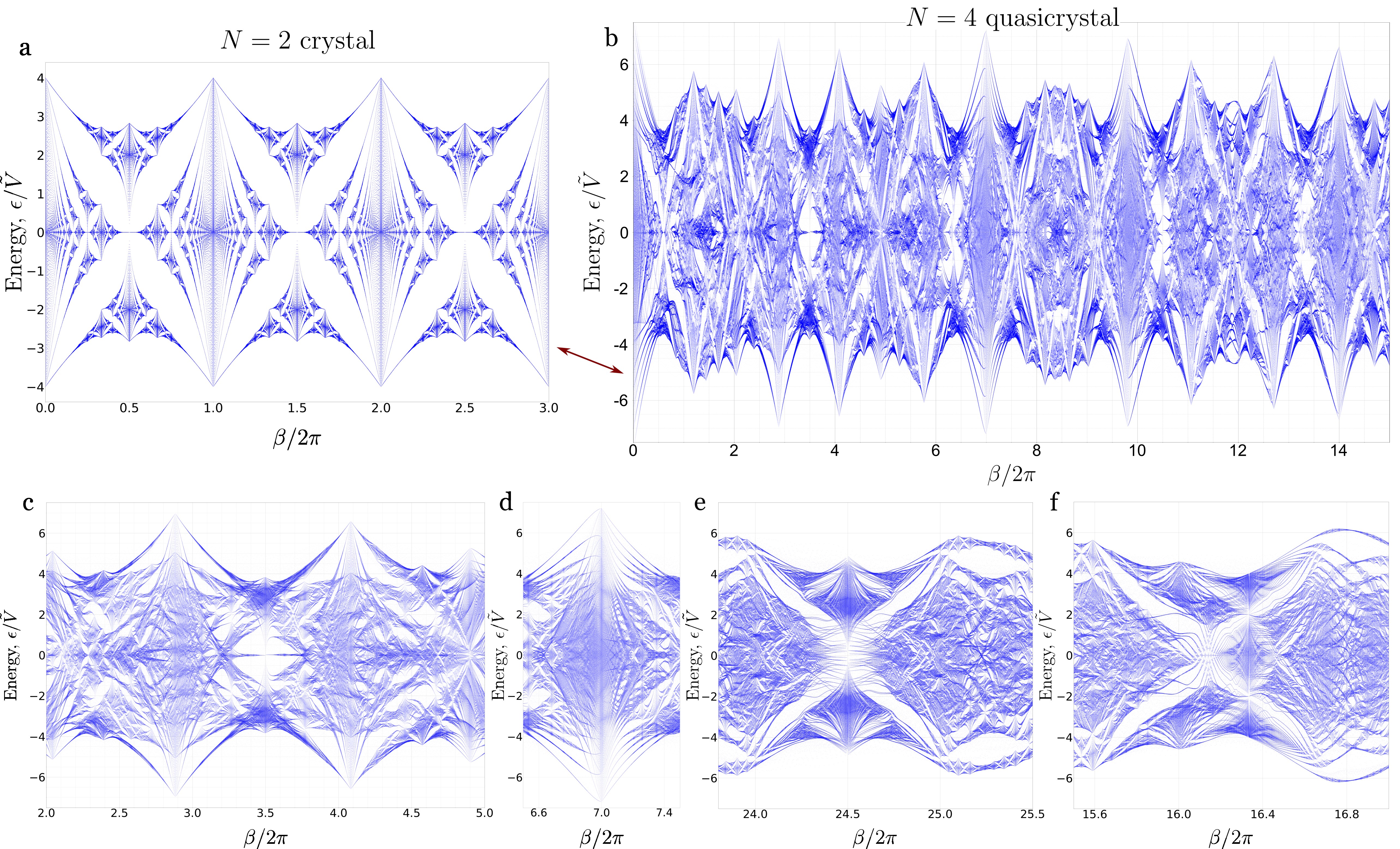} 
    \caption{a), b) A comparison of the 
    crystal and quasicrystal problems. Energy spectra of $H_{N=2}$ and $H_{N=4}$ vs. $\beta$, obtained from a 1D Harper-like problem. For $H_{N=4}$ the incommensurate translations $T_{\pm1}$ and $T_{\pm 1/\sqrt{2}}$ are tackled using rational approximant $\frac{1}{\sqrt{2}} \approx \frac{p}{q}=\frac{5}{7}$ (higher-order approximants were checked to yield similar results, see text). The spectrum is the familiar Hofstadter's fractal for $H_{N=2}$  and has a complex multiscale character for $H_{N=4}$. A double arrow between a) and b) highlights similar fans of Landau-like levels. 
    As in Fig.\,\ref{fig:SKB_spectrum_lines}, the factor $\tilde V$, describing the effective potential strength decay with 
    $\beta$, Eq.\,\eqref{eq:Debye_Waller_factor}, is factored out and included in the energy units. 
    Panels c), d), e), and f) illustrate specific motifs at the locations marked in Fig.\,\ref{fig:SKB_spectrum_lines}. In c), d) the spectrum displays pairs of pyramids filled with fans of Landau-like energy levels, multiple band gaps and complex internal structure. In e), f) large butterfly-shaped bandgaps nested with smaller butterfly-shaped gaps can be seen. Note the web-like patterns discernible at the center of the large butterflies and a fractal web near the right edge. 
    }
    \label{fig:SKB_spectrum}
    \vspace{-5mm}
\end{figure*}

Here we consider Landau levels perturbed by 
octagonal and dodecagonal quasicrystal potentials, Eq.\,\eqref{eq:H_incommensurate}, finding spectra with 
highly unusual 
properties illustrated in Figs. \ref{fig:SKB_spectrum_lines} and \ref{fig:SKB_spectrum} for octagonal quasicrystals. 
The dodecagonal case is overall similar, and is described in \cite{supplement}. While the spectra are clearly far more intricate 
than the Hofstadter butterfly fractal, they share with it common features such as 
fans of Landau-like energy levels and spectral gaps, some of which resemble butterflies. 
However, the most 
striking behavior seen in Fig. \ref{fig:SKB_spectrum_lines} 
is strong oscillations in the spectrum width occurring over a wide range of $\beta$.
The periodicity of these oscillations is incommensurate with the periods expected for the Aharonov-Bohm (AB) or Brown-Zak (BZ) oscillations. 

As we will see, these oscillations reflect a surprising behavior originating from 
non-crystallographic symmetry of the 
quasicrystal problem. Namely, at certain $B$ values the problem acquires a hidden `inner' symmetry, wherein the two square-lattice Hamiltonians comprising Hofstadter's quasicrystal commute.  
This happens at an infinite family of `magic' field values highlighted by red lines in Fig.\,\ref{fig:SKB_spectrum_lines}, forming a $1/B$-periodic family with an irrational period:  
\begin{align}\label{eq:irrational_betas}
B_m a^2= \frac{\phi_0}{\sqrt{2} m}
,\quad m=\pm 1,\pm 2,\pm 3...
\end{align}
where $\phi_0=h/e$ is the flux quantum and $a=2\pi/\kappa$ is the periodicity of the cosine modulation in Eq. \eqref{eq:H_incommensurate}. 
This counterintuitive behavior will be explored below. 

The problem of a particle in a periodic potential, Eq.\,\eqref{eq:H_general}, describes several distinct regimes in which fractal spectra can emerge. 
The best-studied regime is 
a tight-binding problem on a periodic lattice in direct space \cite{Hofstadter_1976,Harper_1955,Azbel_1964}. In this case, the AB interference leads to a periodic dependence on magnetic flux through the lattice unit cell,  $\omega=Ba^2/\phi_0$, where $a^2$ is the unit cell area, with Bloch states disappearing at irrational values $\omega$ and recurring at rational $\omega$, producing Hofstadter fractal. 

A less widely known occurrence of fractal spectra 
is when the cyclotron energy greatly exceeds the periodic potential amplitude. In this case, oscillations in the spectrum, known as the BZ oscillations \cite{BrownZak_Manchester_2017,BrownZak_Manchester_2018}, 
feature periodicity in the inverse magnetic field. 
The relation to the Hofstadter problem can be illustrated for a Landau level perturbed by a weak periodic potential of the form \cite{MacDonald1983} 
\begin{align}\label{eq:H_2cos}
H_{N=2}=2\tilde V\cos \kappa x+2\tilde V\cos \kappa y
,\quad
[x,y]=i\frac{\hbar}{eB}
,    
\end{align} 
where $x$ and $y$ are operators of the electron orbit center $\vec r_c=(x,y)$ and $\kappa $ describes the modulation periodicity. The commutator in Eq.\,\eqref{eq:H_2cos} follows from the relation between electron coordinates and the orbit center radius vector, $\vec r_c=\vec r-\vec v\times \vec z/\omega_c$, $\omega_c=eB/m$. The effective  potential strength $\tilde V$ seen by an electron equals the bare potential strength 
reduced by a ``Debye-Waller factor'' reflecting averaging of $V(\vec r)$ over  electron orbitals. For $n=0$ Landau level this factor equals \cite{MacDonald1983,Goerbig2009,Girvin1999}
\be \label{eq:Debye_Waller_factor}
\tilde V=Ve^{-\kappa^2\ell_B^2/2}=Ve^{-\beta/2}
,\quad
\beta =\kappa^2\ell_{B}^2=2\pi \frac{\phi_0}{B a^2}
,
\ee
with the magnetic length $\ell_B=\sqrt{\phi_0/2\pi B}$ defining the characteristic radius of electron orbitals. This behavior was well understood and documented 
in the early work on fractal spectra in nanostructures \cite{MacDonald1983,Gerhardts1991,Albrecht_2001}.

As a brief detour, we solve the problem in Eq.\,\eqref{eq:H_2cos} by a method that reveals its relation with the Hofstadter problem and explains periodicity in $1/B$. 
This is done by transforming it to Harper's equation, 
a discrete one-dimensional wave equation that can be tackled numerically or analytically \cite{Harper_1955}. 
As a first step, we nondimensionalize the problem by using the magnetic length.  
By rescaling coordinates as $x\to x/\ell_B$, $y\to y/\ell_B$, we absorb $\ell_B$ into a dimensionless wavenumber 
$\tilde \kappa=\kappa \ell_B$. 

Next, using the commutator $[x,y]=i$, we identify $y$ with $-i\p_x$ 
and introduce the one-dimensional translation operators $T_\lambda f(x)=e^{-\lambda\p_x}f(x)=f(x-\lambda)$. This transforms Eq.\,\eqref{eq:H_2cos} 
to a 1D problem $H=2\tilde V\cos\tilde\kappa x+\tilde VT_{\tilde\kappa}+\tilde VT_{-\tilde\kappa}$. After rescaling $x\to \tilde\kappa x$, translation operators become $T_{\pm 1}$,  leading to the Harper problem
\be\label{eq:H_Harper}
H=2\tilde V\cos\beta x+\tilde V T_{+1}+ \tilde V T_{-1}.
\ee
The spectrum dependence on $\beta$, shown in Fig.\,\ref{fig:SKB_spectrum} a), is nothing but the Hofstadter's fractal, periodic under $\beta\to\beta\pm 2\pi$. Periodicity in the inverse $B$ field is in agreement with the properties of the BZ oscillations in a periodic potential observed in strong fields 
\cite{BrownZak_Manchester_2017,BrownZak_Manchester_2018}.

Here, we extend this approach to describe the new regimes arising due to the quasiperiodic potential. We will focus on the $N=4$ quasicrystal case and consider a pair of identical square-lattice potentials given in Eq.\,\eqref{eq:H_2cos} rotated by $45^\circ{}$. This gives a Hamiltonian 
\begin{align}
    H_{N=4}=H_{N=2}+H^{45^\circ}_{N=2}
    ,
    \label{eq:H0+H1}
\end{align}
with the two terms given, respectively, by Eq.\,\eqref{eq:H_2cos} and 
\begin{align}\label{eq:H2}
    H^{45^\circ}_{N=2}=2\tilde V\cos\frac{\kappa (x+y)}{\sqrt{2}}+2\tilde V\cos \frac{\kappa (x-y)}{\sqrt{2}}
\end{align}
Upon rescaling $x$ and $y$ by $\ell_B$ as above, 
$H_{N=4}$ preserves its form and maintains the $8$-fold rotation symmetry, 
whereas the commutation relations become $[x,y]=i$. 

To understand the origin of the `magic' magnetic field values in Eq.\,\eqref{eq:irrational_betas} we consider symmetries of the problem. To that end, we inspect the commutator 
$[H_{N=2}, H^{45^\circ}_{N=2}]$. Its properties can be understood from the commutators of individual cosine terms in Eq.\,\eqref{eq:H_incommensurate}:
\begin{align}\label{eq:H1H2_commutator}
    & [\cos\vec k_1\vec r,\cos\vec k_2\vec r]=\sum_{s_{1,2}=\pm 1}\frac14 
    [e^{is_1\vec k_1\vec r},e^{is_2\vec k_2\vec r}]
    \\ \nonumber
    &=\sum_{s_{1,2}=\pm 1}\frac{i}2\sin\lp\frac{s_1s_2}2(\vec k_1\times\vec k_2)_z\rp e^{is_1\vec k_1\vec r+is_2\vec k_2\vec r}
\end{align}
obtained 
using the Baker-Campbell-Hausdorff 
formula $e^Z= e^X e^Y$, $Z=X+Y+\frac12[X,Y]+...$, where the high-order terms vanish when $[X,Y]$ is a c-number.

For $N=4$ quasicrystals, possible angles between $\vec k_1$ and $\vec k_2$ are $\pm 45^\circ$ and $\pm 135^\circ$ degrees, whereas $|\vec k_{1,2}|=\kappa\ell_B=\sqrt{\beta}$. Therefore, the commutator $[H_{N=2}, H^{45^\circ}_{N=2}]$ vanishes whenever $\beta \frac1{2\sqrt{2}}$ equals $\pi$ times an integer, which gives the condition in 
Eq.\,\eqref{eq:irrational_betas}. 
At such $\beta$'s, the spectrum is a sum of two independent Hofstadter spectra. Away from these values the Hamiltonians $H_{N=2}$ and $H^{45^\circ}_{N=2}$ do not commute. As a result, their eigenstates are hybridized and eigenvalues are reduced. 
Upon approaching the next magic $\beta$ value the eigenstates are unhybridized and eigenvalues grow, giving rise to oscillations. We stress that this behavior, which follows solely from the properties of magnetic translations, is independent of the Landau level number and is therefore expected to be robust, persisting down to the smallest 
$B$ fields. 


                    
Next, we discuss the approach used to obtain the spectrum. 
Following the steps that have led to Eq.\,\eqref{eq:H_Harper}, we arrive at a generalized 1D Harper-like problem involving incommensurate translations by $\pm1$ and $\pm1/\sqrt{2}$
\begin{align}\label{eq:H_4cos}
    &H_4=H_2+H_2^{45^{\circ}}=2\tilde V\cos\beta x+\tilde VT_{-1}+\tilde VT_{+1}
    \\ \nonumber
    & +2\tilde V\cos\lp\frac{\beta}{4}+\frac{\beta x}{\sqrt{2}} \rp 
    T_{\frac1{\sqrt{2}}}
    + 2\tilde V\cos\lp\frac{\beta}{4}-\frac{\beta x}{\sqrt{2}} \rp 
    T_{-\frac1{\sqrt{2}}}
    .
\end{align}
A problem involving incommensurate translations cannot be reduced, per se, to a problem on a periodic lattice. 
To transform it into a Harper-like problem, we use rational approximants of  $1/\sqrt{2}\approx p/q$ found from a step-by-step expansion of $1/\sqrt{2}$ into an infinite continued fraction:
\begin{align}
   \frac1{\sqrt{2}}=\frac1{1+\frac1{2+\frac1{2+\frac1{2+...}}}}
   \to \frac{p}{q}= \frac23,\ \frac57,\ \frac{12}{17} ... \ .
\end{align}
As is well known, truncating infinite continued fractions representing irrational numbers 
yields the best rational approximants with not-too-large denominators. E.g., already $5/7$ approximates $1/\sqrt{2}$ with 
$1\%$ accuracy, which proves sufficient for our purposes.
The next approximant, $12/17$, yields results that support this conclusion.

A rational approximant 
$1/\sqrt{2}\approx p/q$ yields a Hamiltonian with commensurate translations $T_{\pm 1}$, $T_{\pm p/q}$. This is nothing but a 1D tight-binding problem on a lattice with periodicity $1/q$ and non-nearest-neighbor hoppings, a problem that can be readily solved. It is convenient to move to an integer lattice by replacing $x \rightarrow x/q$. 
Upon factoring $\tilde V$, this gives a hermitian Harper-like operator 
    \begin{align}
        \label{eq:H1+2_rat}
       & H_{p/q}= 2\cos\frac{\beta x}{q} + T_{q} + T_{-q}  
       \\ \nonumber 
       & 
       + 2\cos \left(\frac{\beta px}{{q^2}} +\frac{\beta}{4} \right)T_{p}+   2\cos \left(\frac{\beta px}{q^2} - \frac{\beta}{4}  \right)T_{-p} 
       .
    \end{align}
Analyzing the approximant Hamiltonian $H_{p/q}$ starts with defining a Hilbert space. In that, one must exercise caution in choosing the system length and selecting suitable parameter values. First, we consider $H_{p/q}$ on an integer lattice $\{x\}$ using as a basis the Kronecker delta functions, $\braket{x|n} = \delta_{x,n}$, which is an infinite-dimensional Hilbert space. To render it finite-dimensional, ${\rm dim}({\cal{V}})=M<\infty$, we focus on the states $\ket{n}$, $1\le n\le M$ 
and apply periodic boundary conditions $\psi(x+M)=\psi(x)$.

The periodicity requirement, applied to the cosine terms in Eq.\,\eqref{eq:H1+2_rat}, 
restricts possible $\beta$ values. For the Harper problem, a periodicity condition for the cosine term generates a constraint for the allowed $\beta$ values. Here 
we have two cosine terms with different periods, for which periodicity conditions generate two constraints:
\begin{equation}
        {\rm (i)}\quad \beta = 2\pi q k/M; 
        \qquad
        {\rm (ii)}\quad \beta = 2\pi q^2 k'/(p M), 
\end{equation}
with integer $k$ and $k'$. Combined together, these conditions are more stringent than in the Harper problem case, restricting allowed $\beta$ values to
\begin{equation}
    \beta_m = 2\pi m q^2/M 
\end{equation}
with integer $m$.
This gives the minimal allowed step $\Delta \beta = 2\pi q^2/M$ which is $q^2$ times larger than the minimal step found in a similar manner for the Harper problem. 
Which implies that for a fixed step in $\beta$, if we aim to enhance the accuracy of approximating $1/\sqrt{2}$ by increasing the value of $q$, we must also expand the Hilbert space dimension by increasing $M$. In practice, this can prove costly. A better sampling of the spectrum vs. $\beta$ can be achieved by relaxing the constraints imposed by periodic boundary conditions. This yields a spectrum that exhibits a certain level of `noise', which can be traced to the appearance of states localized at the boundary. The resulting spectrum, which is depicted in Figs.~\ref{fig:SKB_spectrum_lines} and \ref{fig:SKB_spectrum},
was obtained by optimizing $\Delta\beta$ in order to achieve high resolution and, at the same time, minimize the added `noise' (parameters used: $M=1960$, $\Delta\beta = 2\pi \cdot 10^{-3}$).

The resulting spectrum, shown in Fig.\,\ref{fig:SKB_spectrum} alongside the Hofstadter butterfly (HB) spectrum obtained by the same method, exhibits pyramid shapes filled with fans of Landau-like energy levels, and a multiscale hierarchy of energy gaps. 
Both the large-scale structure and the fine structure seen in this spectrum differ significantly from those of HB. Nevertheless, closer inspection reveals fragments resembling 
HB-like patterns, as  illustrated in the zoom-ins 
shown in Fig.\,\ref{fig:SKB_spectrum} c), d), e) and f).

However, the most prominent aspect unique to this problem is oscillations vs. $\beta$ with an irrational period. 
It is a signature due to a hidden inner symmetry appearing at isolated $\beta$ values, in general forming an irrational periodic sequence, 
which has no known counterpart in the HB problem and its generalizations [see Fig.\,\ref{fig:SKB_spectrum} a)]. Inspecting 
the dodecagonal problem $H_{N=6}$ \cite{supplement}, confirms the general character of this behavior. 
Occurring at large scales and having a large amplitude, these irrational-period oscillations, like other BZ-type effects \cite{BrownZak_Manchester_2017, BrownZak_Manchester_2018,deVries2023}, are expected to persist to high temperatures and be readily observable in transport or magnetometry measurements. 
The rich and complex structures seen in the spectra 
point to 
a variety of interesting questions, from Brown-Zak band formation and quantum walks in quasicrystal potentials to moir\'e quasicrystal Quantum Hall effect.

We are grateful to Joseph Avron, Leonid Glazman, Philip Kim and Klaus Ensslin for useful discussions. This work 
was performed in part at Aspen Center for Physics, which is supported by 
NSF grant PHY-2210452.

 \section{Supplementary Information 
}

Here we briefly discuss an extension of the results described in the main text to dodecagonal moir\'e quasicrystals. Moir\'e materials with this structure are readily available as they can be produced from pairs of 
layers with a hexagonal lattice structure by a $30^\circ$ twist. 
While being an integral part of state-of-the-art moir\'e toolbox, dodecagonal moir\'e quasicrystals share the key properties established in the main text for octagonal quasicrystals. 

We start by inspecting the Hofstadter problem for electrons on Landau levels in a 2D plane $\vec r=(x,y)$ perturbed by periodic cosine potentials with six-fold symmetry,
\be\label{eq:H_N=3}
V(\vec r)=\sum_{i=1,2,3} V\cos \vec k_i\vec r
,\quad [x,y]=i\frac{\hbar}{eB}
\ee
where $\vec k_1=\kappa (1,0)$, $\vec k_{2,3}=\kappa (-\frac12,\pm \frac{\sqrt{3}}2)$.

\begin{figure}[t]
    \centering
    \includegraphics[scale=1.545]{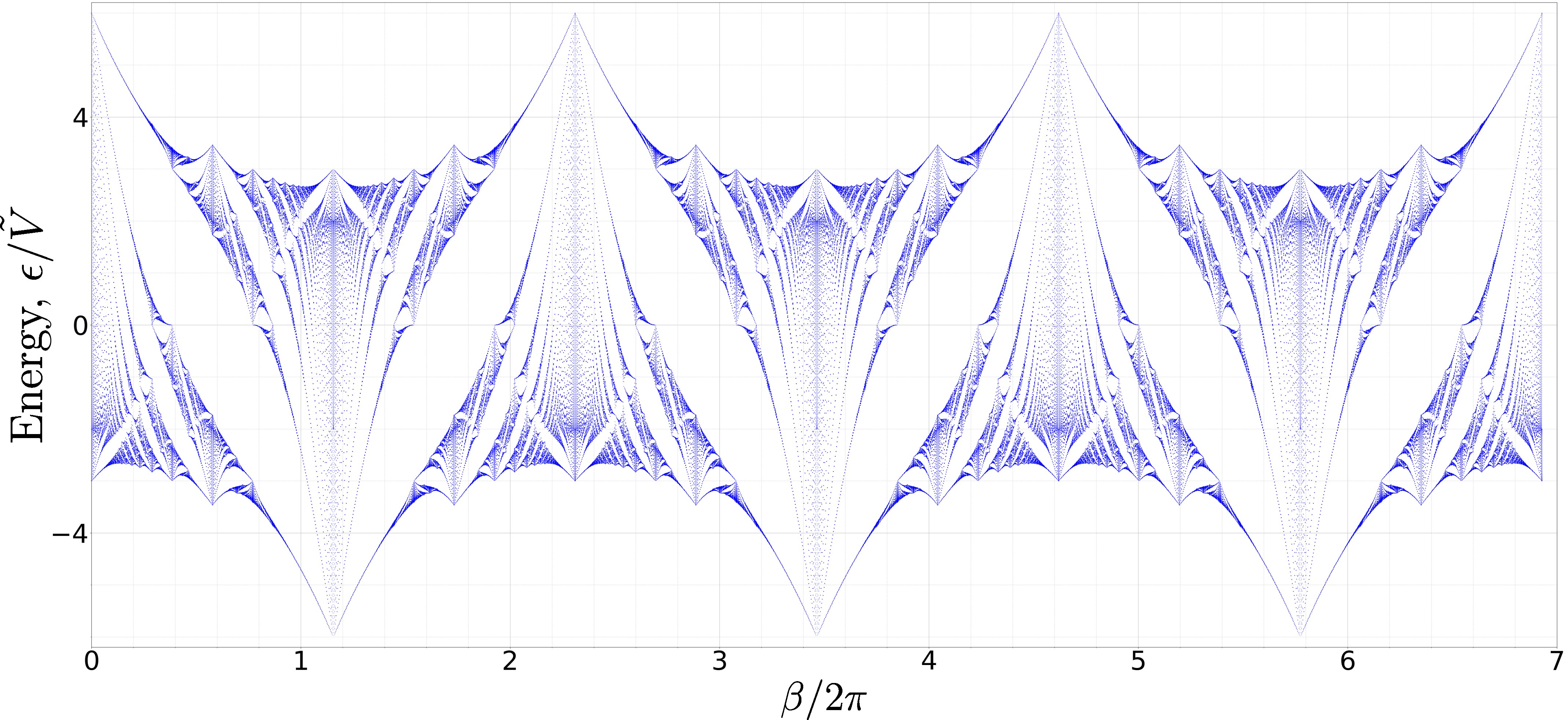}
    \caption{Energy spectrum for a periodic potential with hexagonal symmetry, $H_{N=3}$, vs. $\beta$. The spectrum features periodicity in $\beta$ with 
    with the period $\Delta \beta=4/\sqrt{3}$ and Hofstadter-like fractal patterns identical in each period.}
    \label{fig:HexB_2}
\end{figure}

Initial steps of the analysis follow those in the main text and are totally uneventful. We first rescale all distances by the magnetic length, while preserving the cylindrical symmetry, $(x,y)\to (x/\ell_B,y/\ell_B)$, $\kappa\to\tilde \kappa=\ell_B\kappa$. Then we pass to a 1D representation,  wherein the rescaled commutation relations $[x,y]=i$ yield the relation $y=-i\partial_x$. 
The 1D representation can be brought to a Harper-like form by rescaling $x$ and $y$ by $\tilde \kappa$ and  $\tilde \kappa^{-1}$, respectively, as in the main text, and introducing the translation operators $T_\lambda=e^{-\lambda\partial_x}$. Upon rescaling $x$ by $\sqrt{3}/2$ and factoring $\tilde{V}$, 
the Hamiltonian becomes 
\begin{align}
&H_{N=3} = 2\tilde{V}\cos{\left( \frac{\sqrt{3}}{2} \beta x \right)} + \tilde{V}\cos{\left[\frac{\sqrt{3}}{2} \beta \left(\frac{x}{2} - \frac{1}{4} \right) \right]} T_{1} 
\nonumber \\
&+ \tilde{V} T_{-1} \cos{\left[\frac{\sqrt{3}}{2} \beta \left(\frac{x}{2} - \frac{1}{4} \right) \right]} 
\end{align}
This Hamiltonian yields a `hexagonal variety' of Hofstadter's fractal spectrum, 
which is  
shown in Fig.~\ref{fig:HexB_2}. 
The spectrum features periodic oscillations with a period of $\Delta\beta=4/\sqrt{3}$ and a hierarchical fractal structure within each period. As in the Hofstadter butterfly case, at a generic $\beta$ the spectrum has a structure of a Cantor set of measure zero, whereas at $\beta=\frac2{\sqrt{3}}$ times a rational number the spectrum features Brown-Zak minibands.    

Next, we consider two $N=3$ crystals, described by Eq.\eqref{eq:H_N=3}, twisted relative to each other at a $30^{\circ{}}$ angle and forming a $N=6$ quasicrystal. Similar to the $N=4$ quasicrystal problem solved in the main text, the Hamiltonian involves incommensurate translations ($T_{\pm 1}$, $T_{\pm \frac12}$, and $T_{\pm \sqrt{3}/2}$). Employing a rational approximation 
\be
\sqrt{3}\approx p/q
\ee 
yields a 1D tight-binding problem with non-nearest-neighbor hopping matrix elements defined on a periodic 1D lattice. The resulting Harper-like operator describing $N=6$ quasicrystal is 
\begin{align}
     & H_6 = \tilde V T_{2q} +\tilde V T_{-2q} + 2\tilde V\cos{\left(\frac{\beta x}{2q}\right)} 
     \\ \nonumber  &+\tilde V \cos{\left(\frac{\beta x p}{4q^2} - \frac{\beta p}{8q}\right)} T_{q} 
     + \tilde V T_{-q} \cos{\left(\frac{\beta x p}{4q^2} - \frac{\beta p}{8q}\right)} 
     \\ \nonumber
     &
     +\tilde V \cos{\left(\frac{\beta x}{4q}  - \frac{\beta p}{8q}\right)} T_{p} 
     +\tilde V T_{-p}\cos{\left(\frac{\beta x}{4q}  - \frac{\beta p}{8q}\right)} 
     . 
\end{align}
Numerical diagonalization is carried out using the same Hilbert space as for the $N=4$ quasicrystal. Namely, we employ a finite-dimensional Kronecker delta-function basis, $\braket{x|n} = \delta_{x,n}$, where $n$ takes the values $\ket{n}$, $1\le n\le M$.  
Diagonalizing a $M\times M$ matrix yields the spectrum of the problem vs. $\beta$. 

\begin{figure*}[t]
    \centering
    \includegraphics[scale=1.6845]{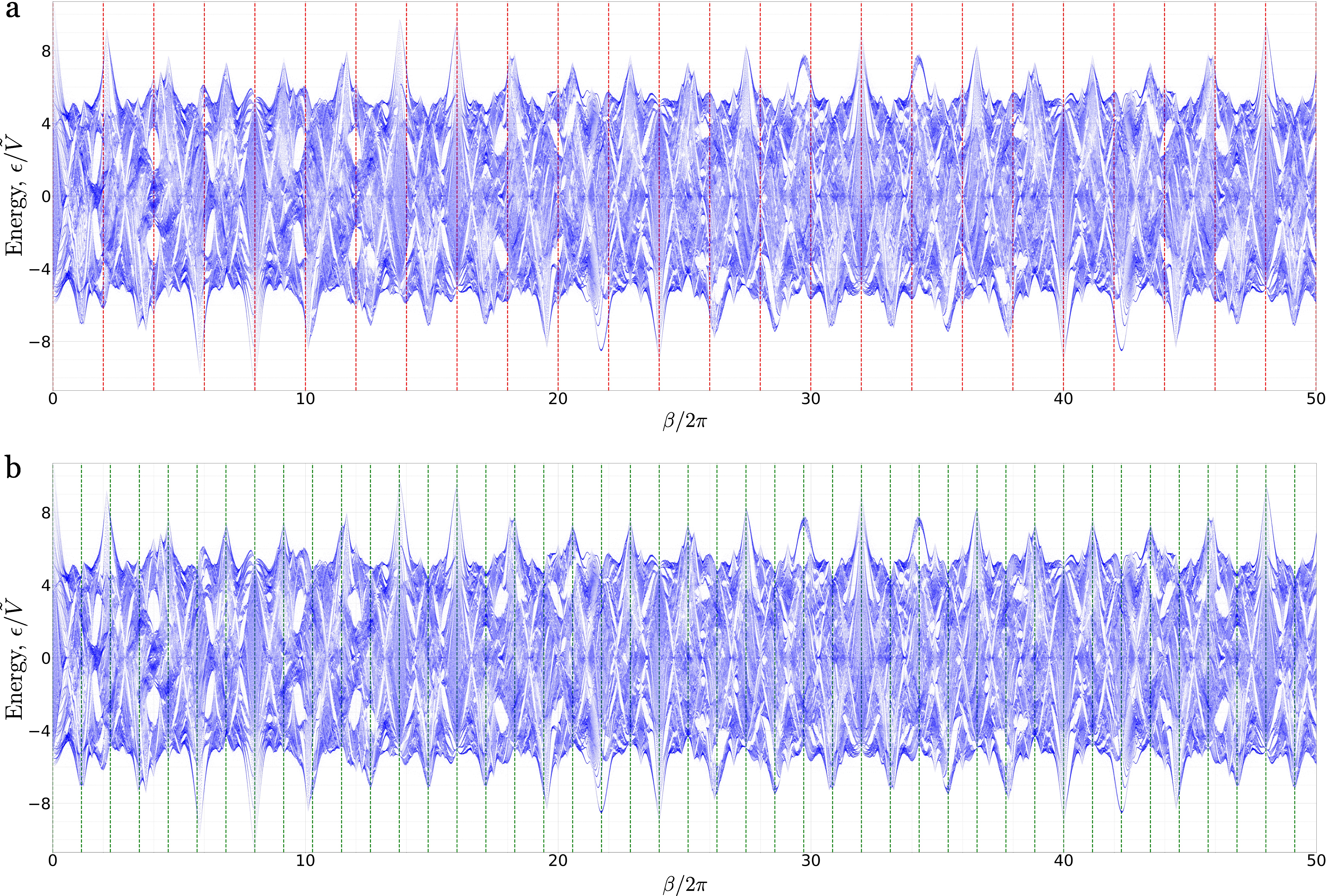}
    \caption{Energy spectrum of a quasicrystal $H_{N=6}$ vs. $\beta$ obtained from a Harper-like 1D problem (see text). 
    The spectrum features approximately periodic sequence of positive and negative pyramids filled with Landau-like fans of energy states as for a single hexagonal crystal problem [see Fig.\ref{fig:2HexB_2MP}]. a) The red lines 
    mark the hidden symmetry points where the two hexagonal crystal Hamiltonians commute (Eq.~\ref{eq:Hex_commutation_of_Hamiltonian}). b) The green lines 
    mark $\beta$ values for which 
    different terms of the periodic crystal Hamiltonians commute (Eq.~\ref{eq:Hex_commutation_of_terms}).}
    \label{fig:2HexB_2MP}
\end{figure*}

Fig.~\ref{fig:2HexB_2MP} shows the spectrum obtained from the rational approximant $\sqrt{3} \approx 7/4$ and system size $M=784$. While the quasicrystal spectrum 
is far more complex than that for the $N=3$ crystal problem, 
some familiar patterns emerge, such as fans of Landau-like energy levels and a multiscale hierarchy of the forbidden and allowed energy bands. 
Higher-order approximant $p/q=19/11$ was checked to yield 
similar results. 

Next, we proceed to identify 
the `magic' $\beta$ values for which the quasicrystal Hamiltonian displays hidden symmetry. As we will see, here the pattern of hidden symmetries is somewhat different than that for the octagonal quasicrystal case. 
First, 
we find $\beta$ values for which the Hamiltonians of two underlying hexagonal lattices commute. Eq.~\eqref{eq:H1H2_commutator} yields the following two conditions:
\begin{equation}
       1)\ \  \frac{\beta}{2\pi} \sin{\left( \frac{\pi}{6} \right)} = n, \qquad 
        2)\ \  \frac{\beta}{2\pi} \sin{\left( \frac{\pi}{2} \right)} = m,
\end{equation}
with integer $n$ and $m$. These conditions give $\beta$ values 
\begin{equation}\label{eq:Hex_commutation_of_Hamiltonian}
    \frac{\beta}{2\pi} = 2n 
\end{equation}
with integer $n$. 
These points are depicted by red lines in Fig.~\ref{fig:2HexB_2MP} a). Notably, similar to the octagonal case discussed in the main text, the periodicity of magic points differs from the periodicity 
in the spectrum of the $N=3$ crystal by an irrational factor. 

In addition, the $N = 6$ quasicrystal problem also features another set of magic $\beta$ values, associated with what is appropriate to call `quasi-symmetries'. Quasi-symmetries arise as follows. 
Among the six wavevectors $\bm{k}_s$ (Eq.~\ref{eq:H_incommensurate}) defining this quasicrystal, the angles between three 
wavevectors in the set $K_a=\{ \bm{k}_1, \bm{k}_3, \bm{k}_5 \}$ 
are $\pi/3$ or $2\pi/3$. Therefore, the commutators $[\cos \vec k_s\vec r, \cos \vec k_{s'}\vec r]$ vanish for the same $\beta$ values.  Terms that correspond to vectors from $K_a$ pairwise commute provided
$\frac{\beta}{2\pi} \sin\lp \frac{\pi}{3} \rp = n$
with integer $n$. This condition yields $\beta $ values  
\begin{equation}\label{eq:Hex_commutation_of_terms}
    \frac{\beta}{2\pi} =  \frac{2}{\sqrt{3}}n
    . 
\end{equation}
The same 
is true for the other three wavevectors forming the set $K_b=\{ \bm{k}_2, \bm{k}_4, \bm{k}_6 \}$.
Therefore, at the $\beta$ values given in Eq.\eqref{eq:Hex_commutation_of_terms}, the terms of the Hamiltonian corresponding to the vectors from $K_b$ also pairwise commute. These points are shown in Fig.~\ref{fig:2HexB_2MP} b) by green lines. The periodicity of this $\beta$ sequence is the same as the period of oscillations in the spectrum of $N=3$ crystal. Clearly, 
the green lines describe periodicity 
in the spectrum vs. $\beta$ better than the red lines. 

Therefore, in direct analogy to the octagonal case, the problem exhibits strong nearly periodic oscillations in the spectrum and  a hidden `inner' symmetry arising at $\beta$ values for which the $K_a$ and $K_b$ parts of the Hamiltonian pairwise commute. Yet, perhaps surprisingly, the periodicity in the spectrum is explained by `quasi-symmetries' occurring when the terms within the $K_a$-only and $K_b$-only parts of the Hamiltonian, taken individually, pairwise commute. The role of quasi-symmetries in this problem is an interesting topic for future work.  



\end{document}